# Resistive switching of tetraindolyl derivative in ultrathin films: A potential candidate for non-volatile memory applications

Surajit Sarkar[†], Hritinava Banik[†], Sudip Suklabaidya[†], Barnali Deb[§], Swapan Majumdar[§], Pabitra Kumar Paul[∥], Debajyoti Bhattacharjee[†] and Syed Arshad Hussain*[†]

[†]Thin Film and Nanoscience Laboratory, Department of Physics, Suryamaninagar, Agartala 799022, West Tripura, Tripura, India

[§]Department of Chemistry, Tripura University, Suryamaninagar, Agartala 799022, West Tripura, Tripura, India

[∥]Department of Physics, Jadavpur University, Jadavpur, Kolkata 700032, India

* Corresponding author

Email: sa_h153@hotmail.com, sahussain@tripurauniv.in (SAH)

Ph: +919402122510 (M), Fax: +913812374802 (O)

**Abstract.** Bipolar resistive switching using organic molecule is very promising for memory application owing to their advantages like simple device structure, low manufacturing cost, theirstability and flexibility etc. Herein we report Langmuir-Blodgett (LB) and spin-coated film based bipolar resistive switching devices using organic material 1,4-bis(di(1H-indol-3-yl)methyl)benzene (**1**). Pressure - area per molecule isotherm ($\pi$ – A), Brewster Angle Microscopy (BAM), Atomic Force Microscopy (AFM) and Scanning Electron Microscopy (SEM) were used to have an idea about organization and morphology of the organic material onto thin film. Based on device structure and measurement protocol it is observed that the device made up of **1** shows non-volatile Resistive Random Access Memory (RRAM) behaviour with very high memory window (~$10^6$), data sustainability (5400 sec) and repeatability. Oxidation-reduction process as well as electric field driven conduction are the key behind such switching behaviour. Due to very good data retention, repeatability, stability and high device yield the switching device designed using compound **1** may be a potential candidate for memory applications.

**Introduction**

Resistive switching devices using organic molecules are of gaining interest in the area of memory application due to the added advantage of using organic molecules.[1–7] At present majority of the memory devices available are based on silicon.[8,9] However, devices made using silicon basedmaterial has some limitations such as low write-erase speed, limited reusability of flash memory, the volatility of Dynamic Random Access Memory (DRAMs), slow response of Hard Disk Drives (HDDs) in the magnetic field trigger the need for alternative memory materials. In this regard memory devices using organic materials can substantially overcome the said limitations and therefore memory applications using suitable organic materials are very attractive.[10] Resistive switching suitable for Resistive RRAM device based on organic materials showing non-volatile bipolar switches may be a potential replacement for silicon in the memory industry. There are several advantages of these organic materials as memory element such as – very good read-write speed, scalability, easy device fabrication, excellent data endurance, retention, repeatability, very low power requirement and good compatibility with existing CMOS technology.[11–16]

Apart from memory application, RRAM can be used in future artificial intelligence and neuromorphic computing.[17–20] Metal-insulator-ITO based RRAM has organic material sandwiched between two electrodes. It can be switched between two resistance states- low resistance state (LRS) and high resistance state (HRS) based on bias voltage (from high to low or from low to high). Depending upon the data retention characteristics of bipolar (non-volatile) switching and threshold switching (volatile) are promising candidates for future memory technology.[21] In bipolar switching, after removal of external voltage both the HRS and LRS are retained, whereas in threshold switching only HRS is retained at a low applied voltage.[21–24]

Of late much effort has been made, in the development and investigations of various new organic materials suitable for opto-electronic applications. Materials preferably polar molecule, which can be easily polarised are suitable for opto-electronic device fabrication. Molecules having π-conjugated system as well as electron donor and/or acceptor group form a push pull system (D-π-A). These types of molecules posses inter molecular charge transfer as well as very low energy barrier. Also different five and six number hetero cycles may act as auxiliary acceptors or donors and can be used as π-conjugated backbones. This may include the overall performance of the device employing such molecules.[25–27] In this context, indole derivatives are very much important heterocyclic systems because of its π-conjugation and presence of –NH group of its pyrrole ring it is expected to have interesting electronic properties.

Recently we have synthesized several indole derivatives and studied their optical behaviour.[28] Interestingly it has been observed that self-standing films of indolyl molecule mixed with clay particle showed optical switching under suitable condition.[28] Due to the oxidation-reduction capability as well as the presence of high π-electron cloud into the molecular backbone of the indolyl molecule we thought that this molecule may have interesting electrical behaviour as well. Accordingly we have studied the resistive switching behaviour of 1,4-bis(di(1H-indol-3-yl)methyl)benzene (**1**) molecule assembled onto LB and Spin-Coated film. Interestingly our results suggest that switching device based on Indole**1** molecule showed bipolar resistive switching behaviour with very high memory window (~$10^6$). This study suggests that indolyl based resistive switching device is very promising for use in next-generation high-performance non-volatile memory applications using organic molecules.

## Experimental Section

### Materials

We have synthesized the organic materials Indolyl derivatives namely 1,4-bis(di(1H-indol-3-yl)methyl)benzene (**1**) and p-Di[3,3'-bis(2-methylindolyl)methane]benzene (**2**) which are reported in our previous work.[28] Terephthaldehyde and DDQ were procured from Spectrochem Pvt. Ltd. and used as received. Working solution was prepared using Methanol (Spectroscopic grade, SRL, India) and Chloroform (Spectroscopic grade, SRL, India) as solvent. Ultra-pure (18.2 MΩ–cm, Milli-Q) water is used as sub-phase.

### Isotherm measurement and LB film formation

Measurement of π - A isotherm and LB film deposition was being done by means of LB film deposition instrument (Apex 2000C, Apex Instruments Co., India). 90 µl of chloroform solution (0.5 mg/mL) of Indole1 was spread using a microsyringe onto the subphase of pure Milli-Q water in order to measure the π - A isotherm as well as for film preparation. Π - A isotherm was recorded at 5 mm/min barrier compression rate, after evaporation of chloroform. The π - A isotherm was measured with the help of Wilhelmy plate arrangement system. For film deposition, silicon-wafer was used for AFM study, quartz slide was used for spectroscopic study and ITO coated glass slide was used for I-V measurement respectively. Details of the LB technique has been reported elsewhere.[29]

### Fabrication of spin-coating film

Spin-coating film deposition instrument (SCU 2005A, Apex Instruments Co., India) available in our lab was used for preparing the film of **1** and **2** onto ITO coated glass slide as well as onto quartz slide. 500 µL chloroform solution of Indole1 and Indole2 (0.5 mg/mL) was spread onto a cleaned quartz glass slide as well as on ITO slide drop-wise followed by spinning such that the sample solution spreads almost uniformly over the whole surface of the slide. The substrate was allowed to spun (Rotational speed was 1500 rpm for 120 seconds) for every 1-2 drop of the sample solution. Details of the spin-coating technique have been reported elsewhere.[30]

### BAM imaging

Morphological structure of the floating film at air-water interface was done by BAM (Accurion nanofilm_EP4-BAM, Serial No. 1601EP4030). The incident light used was p-polarized (30 mW, 532 nm). Details of the BAM imaging has been reported elsewhere.[23]

### AFM imaging

A commercially available AFM (Model: Innova, Bruker AXS Pte Ltd.) was used to take the AFM images of LB films. Intermittent-contact ("tapping") mode was used during imaging. Details of the AFM imaging has been reported elsewhere.[31]

### Cyclic Voltametry (CV)

A commercially available Electrochemical workstation CorrTest was utilized to observe the voltametric response of the Indole1-LB film through Cyclic Voltametry experiment. During the measurement scanning speed was 100 mV/s.

### I - V characteristic measurement

I – V characteristics were investigated using sourcemeter from Keithley (Sl No. 2401). The device configuration was Au/organic layer/ITO. Organic layer was deposited onto ITO coated glass slide using either LB or spin coating technique. For LB film based device 60 layer of organic material were deposited onto ITO coated glass slide. Schematic representation of the device structure is shown in figure S1 of supporting information. LB films were deposited at 30 mN/m surface pressure with a deposition speed 5 mm/min. For all devices, ITO coated glass substrate and gold were used as bottom and top electrode respectively. Array of circular pads of gold with average diameters of the order of ~500 μm were deposited using designed mask. The prepared films were dried in vacuum desiccators for at least 10 hours before observation. I-V characteristics were measured in various sweep direction with step potential 2.5 mV/sec and during the measurement active area was 1 mm$^2$.

## Result andDiscussion

**Floating monolayer at air water interface:**

**Pressure - area (π - A) isotherm:** Preparation of floating film (Langmuir film) at the air-water interface is the first step to prepare LB film onto solid substrate.[29] Accordingly, we studied the thermodynamic behaviour of **1** molecule at the air-water interface by measuring and analyzing the pressure area isotherm.[29] Corresponding isotherm curve is shown in figure 1.

Isotherm curve shows a continuous rise in surface pressure with a decrease in area per molecule with an initial lift-off area 0.745 nm$^2$.[31] The mean molecular area (area per molecule at solid phase) obtained is 0.446 nm$^2$ (figure S2 of supporting information). To have an idea about the orientation of **1** molecule in the Langmuir film at solid phase we have analyzed the cross-sectional area of the molecule and compared the same with mean molecular area.[22] Assuming the Indole1 molecule as a cuboid box having a dimension of three sides as 1.412 nm, 0.904 nm and 0.5 nm, the three cross-sectional areas calculated are 0.452, 0.706 and 1.276 nm$^2$ (figure S3 of supporting information). A close look to these values suggests that mean molecular area (area per molecule at solid phase) is very close to 0.452 nm$^2$ i.e., cross-sectional area corresponds to the vertical orientation of the molecule. This indicates that Indole1 molecule lies vertically oriented within the Langmuir film when the solid phase is reached. This also suggests that Indole1 molecule form one molecule thick packed monolayer film at air-water interface at higher surface pressure.[22] Also, we have calculated the compressibility of the floating film within 5 - 15 mN/m and 25 – 35 mN/m surface pressure region following the procedure described elsewhere.[32] It has been observed that the compressibility value at higher surface pressure region (29.85 mN$^{-1}$) is lower than the lower surface pressure region (40.42 mN$^{-1}$).[23,32] This also supports that at high surface pressure (25 – 35 mN/m) region **1** molecule form more compact film.[23] Almost negligible variation in surface pressure was observed when the barrier was fixed at different surface pressure viz. 25, 30 mN/m etc during isotherm measurement.[23] This also suggests that Indole1 molecules form stable Langmuir film at higher surface pressure. Accordingly, we have chosen the 30 mN/m surface pressure for LB film deposition onto solid support for switching device fabrication and characterization. Our later investigation at BAM, SEM and AFM studies also confirm that at this surface pressure Indole1 molecules form smooth, uniform and almost continuous film onto solid support which is the prerequisite for electronic device applications.[33]

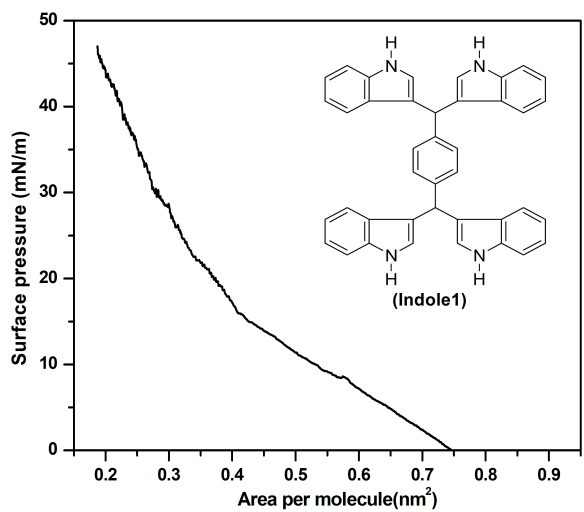

**Figure 1.** π - A isotherm of **1** molecule on water sub-phase. Inset represents the chemical structure of 1,4-bis(di(1H-indol-3-yl)methyl)benzene (**1**) molecule.

**BAM imaging of Langmuir films:** To have a visual idea about the Langmuir film formation of Indole1 molecule at the air-water interface, traditional imaging technique in-situ BAM has been employed. BAM image of empty water surface before spreading of the molecules was completely dark which indicates no reflection of light occurred from the empty surface (figure S4 of supporting information).[23] BAM images of the Langmuir film of **1** measured at different surface pressure are shown in figure 2. Image measured before compression of the barrier showed large circular empty areas with molecules surrounding these areas (figure 2a). Surface coverage was very less. However, with increasing surface pressure by compression surface coverage in the film increases resulting lowering of the empty circular areas in the film (figure 2b). Interestingly at around 30 mN/m surface pressure, Indole1 molecules form a continuous smooth film with almost no empty space (figure 2c). Isotherm studies of the Langmuir film also showed that at around 30 mN/m surface pressure maximum compact uniform film was formed at air-water interface.

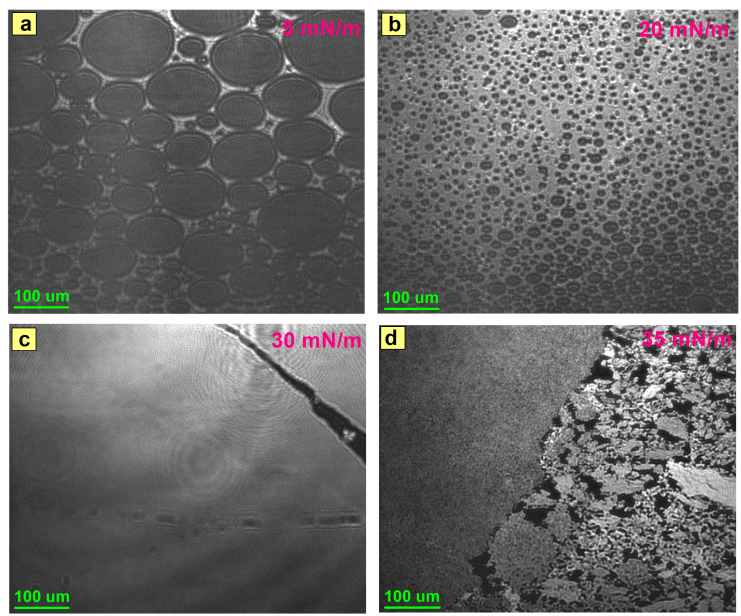

**Figure 2.** BAM images of the Indole1 monolayer recorded during π - A measurement at air-water interface.

For checking the stability of the film at 30 mN/m surface pressure barrier was kept fixed and successive BAM images were recorded up to 5 hours. Also, BAM images were recorded through successive compression-expansion cycles. No significant changes in the corresponding BAM images measured with time and compression-expansion cycle were observed (Figure not shown). These observations indicate that Indole1 molecules form stable Langmuir film at air-water interface at around 30 mN/m surface pressure. This justifies the selection 30 mN/m surface pressure to deposit floating film onto ITO coated glass substrate for switching device fabrication. However, the BAM images recorded beyond 35 mN/m showed significant changes in the film structure. Here continuous film breaks partially; this indicates that beyond 35 mN/m surface pressure partial collapse of Langmuir film may occur at the air-water interface.

**AFM and SEM imaging of LB films:** In LB technique transfer of floating Langmuir film onto a solid substrate is very crucial. Film structure largely depends on the deposition. In some cases, changes in film structure with respect to Langmuir film have been observed.[34] Therefore, to have idea about the film structure, we have investigated LB monolayer films of Indole1 deposited at 30mN/m surface pressure using AFM and SEM. Representative images are shown in figure 3.

AFM images of LB monolayer film showed that Indole1 Langmuir film has been successfully transferred onto a silicon substrate with very good surface coverage (>80%). Height profile analysis shows that film thickness lies within 1.5 to 2 nm range (figure 3b). On the other hand, SEM images (figure 3c) also confirmed the formation of very smooth and continuous LB film. As a whole both AFM and SEM studies give compelling visual evidence of successful deposition of uniform LB film onto silicon wafers at chosen surface pressure.

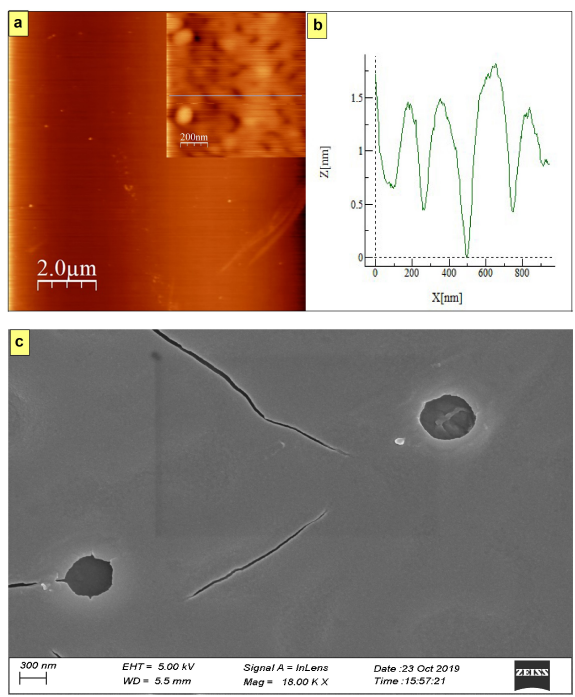

**Figure 3.** (a) AFM image (b) height profile analysis of AFM image and (c) FESEM images of Indole1 molecule in the monolayer LB film deposited at 30 mN/m.

**Resistive Switching Phenomenon:**

Due to the presence of oxidizable and reducible group as well as high π-electron cloud, it is expected that Indole1 molecule might show interesting electrical behaviour under suitable condition. Accordingly, I-V characteristics of Indole1 assembled onto LB film has been investigated. The main interest was to check the resistive switching behaviour using Indole1 molecule. Here LB technique was used to prepare switching device as this technique allows the organization of molecules onto uniform thin-films with molecular-level control.[23,34] LB film based resistive switching device using Indole1 molecules might be suitable for memory application and future organic electronics.[35,36] It may be interesting to mention in this context that we have observed that Indole1 molecule showed interesting optical behaviour assembled onto self-standing films.[28] Other indole derivative with interesting optoelectronic behaviour has also been reported.[37]

In order to study the I-V characteristics, 60 layers LB films of Indole1 molecules have been deposited onto ITO coated glass substrate. The device configuration was Au/Indole1-LB film/ITO. Here Au and ITO acted as top and bottom electrode respectively. While studying the switching behaviour I-V characteristics were recorded by applying scanning voltage in both forward (+$V_{max}$ to -$V_{max}$) and reverse (-$V_{max}$ to +$V_{max}$) sweep direction.

Typical I-V characteristics for the designed device is shown in figure 4 presented in linear scale (figure 4a) and semi-log scale (figure 4b), while measuring from +$V_{max}$ to -$V_{max}$. Arrows in the figure show the scanning direction. At the onset of scanning the device shows very low current and remains at high resistance state (HRS). When the scanning voltage approaches the typical switching threshold of about -2.45 V, the current through the device increased abruptly and the device switched to its ON state i.e., from HRS to low resistance state (LRS). This is called SET operation and the corresponding threshold voltage is known as the SET voltage.

Again when the scanning voltage is reversed i.e., $-V_{max}$ to $+V_{max}$, the device switched back to its HRS state when the scanning voltage is reached at about +2.58 V. At this point, the device is RESET from its ON state to its original OFF state (HRS). Corresponding voltage is called RESET voltage. This type of bipolar switching using organic molecules is very promising for applications such as memory[38], computing[39], sensing[40], logic circuit[41] etc.

In these bipolar switching devices, the SET (transition from HRS to LRS) operation represents a 'Writing' operation in case of memory application.[23,42] The ON state (LRS) can be retained in the device even after removal of bias and the device subsequently returned to its OFF state (HRS) (erasing process) under an opposite bias.[23,42]

In order to check the directional dependence on the observed switching, we have also studied the I-V characteristics by reversing the initial sweep direction. Here initial scanning was applied from $-V_{max}$ to $+V_{max}$. In this case the SET operation i.e., HRS to LRS transition occurred at positive voltage whereas RESET (LRS to HRS) occurred at a negative voltage (figure not shown). But the amplitude of SET and RESET voltage remains almost unchanged with respect to the previous scanning direction. This suggested that in the present case the observed bipolar switching was independent of initial scan direction.

In the present case, the scanning voltage range used was +3 V to -3 V. It was observed that when $V_{max}$ exceeds ±3.5 V, once the device switched to its high conductive ON state, it did not return back to its OFF (HRS) state when the scanning voltage was reversed (figure. 4c and 4d). This may be due to the short circuit of the device.[43] When current passes through the organic layer, the layer may be heated up. Depending on the bias voltage if the current crosses a certain limiting value the organic layer within the device may be damaged due to heating, resulting in a short circuit in the device.[43] So in the present device, the allowed scanning range is ±3.5 V.

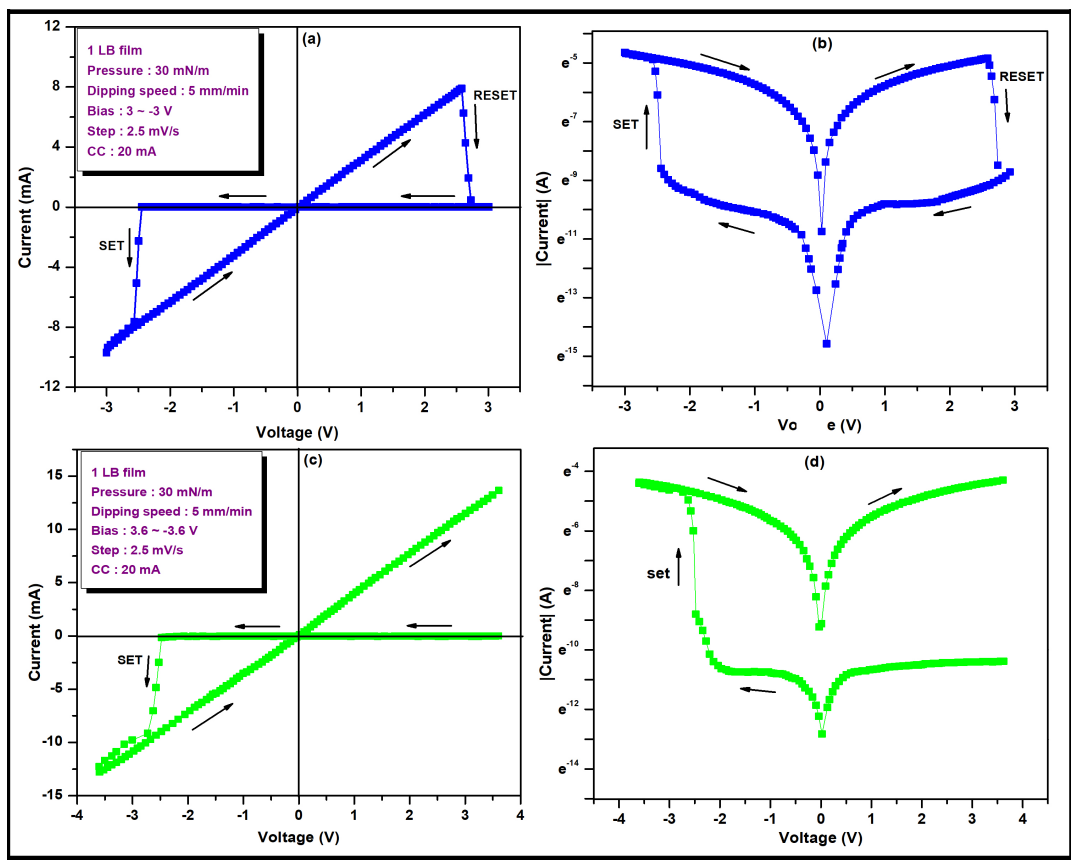

**Figure 4.** I-V measurement showing bipolar resistive switching in linear scale (a & c) and in semi–log scale (b & d) of the Au/Indole1/ITO device based on 60 layer LB films of Indole1 for two sweep direction. Initial sweep voltages were (a) ± 3 V, (c) ± 3.6 V. Arrows represents the direction of sweep depending on the applied voltage.

In order to check the importance of LB technique, during device formation on switching performance, we have also used spin-coated film in between electrodes with configuration Au/Indole**1**-Spin coated film/ITO and observed the I-V characteristics. Corresponding I-V curves are shown in figure S5 of supporting information. Here also similar bipolar switching behaviour was observed with SET and RESET voltages -3.51 V and +3.56 V respectively. In the case of LB films, the switching voltages were much lower. For practical and commercial application lower switching voltage is advantageous.[42]

Various switching parameters as extracted from I-V curves showing bipolar switching for both the above devices along with some other similar reports of switching using organic molecules are listed in Table-1. ON/OFF current ratio is an important criterion for switching devices in case of memory applications.[21] High ON/OFF ratio is always expected for reliable memory performance.[44] Interestingly in the present case both the devices show very high ON/OFF ratio 1.67 x $10^6$ for LB film and 3.8 x $10^4$ for the spin-coated film. A comparison with other reported results also suggested that the observed value of ON/OFF ratio for LB film device (~$10^6$) is quite high (Table 1). Therefore present LB film based switching device may have a very high potential for reliable memory application.[44] However, in the present case, the difference in performance for two devices as observed is mainly may be due to the film preparation technique used. It is well known that the LB technique is very effective for the preparation of smooth, continuous and uniform ultra-thin films compared to spin coating technique.[23] Thus morphology of organic layer during switching device formation may be mainly responsible for the better performance in LB film based devices.

**Table 1.** Performance comparison of Indole1 based memory and other organic material based memory devices.

| Device Configuration | Set and Reset Voltage (V) | | OFF state resistance ($\Omega$) | ON state resistance ($\Omega$) | Current ON/OFF ratio | Retention time(sec) | References |
|---|---|---|---|---|---|---|---|
| | $V_{SET}$ | $V_{RESET}$ | | | | | |
| Au/phenazine/Au | -1.3 | +1.4 | $\sim 5 \times 10^4$ | $\sim 4 \times 10^2$ | >100 | 10000 | 21 |
| Au/Imidazole/ITO | -1.95 | +1.95 | $34.5 \times 10^4$ | $46 \times 10^1$ | 794 | - | 22 |
| Au/MnTPPS/ITO | -4.1 | +4.1 | $10.9 \times 10^4$ | $38 \times 10^1$ | 286 | - | 23 |
| Ag/melanin/SS | +0.6 | -0.6 | $\sim 6 \times 10^5$ | $\sim 5 \times 10^4$ | 10 | 1000 | 42 |
| Ag/CGMFs/FTO | +1.33 | -1.42 | $\sim 920$ | $\sim 210$ | 4.37 | 2000 | 45 |
| Al/silk fbroin/ITO | +14 | -14 | $\sim 3 \times 10^3$ | $\sim 2.7 \times 10^2$ | 11 | 1000 | 46 |
| Ag/pectin/FTO | +3.3 | -4.5 | $\sim 4.5 \times 10^6$ | $\sim 1 \times 10^4$ | 450 | $\sim 900$ | 47 |
| Al/parylene/W | +2.2 | -0.6 | $\sim 5 \times 10^6$ | $\sim 4.5 \times 10^2$ | $10^4$ | 100000 | 48 |
| Ag/sericin/Au | +2.5 | -1 | $\sim 2.2 \times 10^{10}$ | $\sim 4.9 \times 10^3$ | $10^6$ | 1000 | 49 |
| Pt/proton-doped polyazomethine/Pt | +3 | -7 | $\sim 8.2 \times 10^6$ | $\sim 4.6 \times 10^4$ | 10 | - | 50 |
| Au/Indole1-Spin coated film/ITO | -3.51 | 3.56 | $2.041 \times 10^7$ | $5.37 \times 10^2$ | $3.8 \times 10^4$ | 5400 | Present work |
| Au/ Indole1-LB film /ITO | -2.45 | 2.58 | $5.09 \times 10^8$ | $3.04 \times 10^2$ | $1.67 \times 10^6$ | 5400 | Present work |

Data reproducibility and sustainability are very much essential in order to have a reliable practical and commercial application of memory devices. In order to check the data sustainability, we have checked the data retention test for both the devices, this gives an idea about the ability of a memory device to retain the stored data i.e., a particular state (HRS or LRS) for prolonged time.[45] Here we have investigated the data retention characteristics for both the devices by studying the LRS and HRS with time. The LRS and HRS of the designed devices were recorded every 150 seconds over a time period of 5400 seconds for both the devices at reading voltage 0.5 V. A cumulative distribution of OFF and ON state resistances is represented in figure 5. The results suggested that for both the devices, the LRS and HRS values are highly stable over a time period of 5400 seconds with a very high memory window.

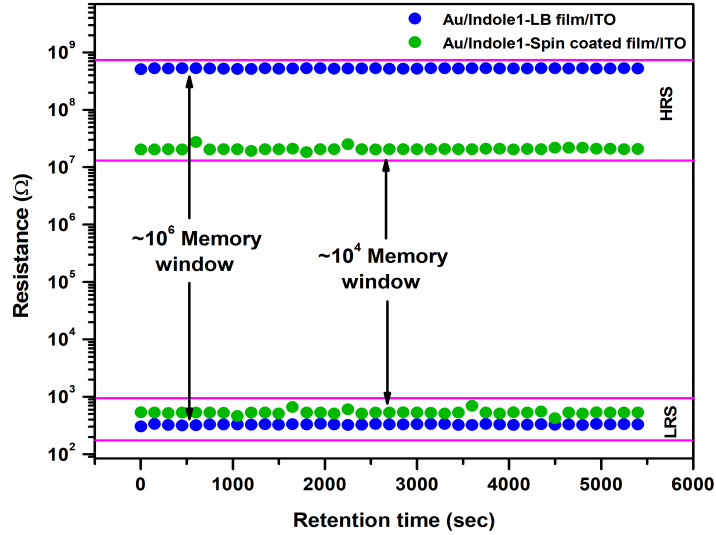

**Figure 5.** Retention characteristics of the Au/Indole**1**/ITO device with LB film (blue colour) and spin coated film (green colour).

In order to check the repeatability of the designed memory device, we have investigated LB film based devices under several consecutive SET and RESET processes. Results suggested that the designed switching device switches between LRS to HRS and vice versa over 50 cycles with stable well resolved two resistance states having memory window $\sim 10^6$. Representative switching curves are shown in figure 6. Inset shows the cumulative distribution of $V_{SET}$ and $V_{RESET}$ voltages. The results showed that the device exhibits a very good uniform and almost reproducible switching behaviour. The average values of $V_{SET}$ and $V_{RESET}$ are -2.418 V and +2.658 V respectively with a standard deviation of 5.35% and 11.89% respectively.

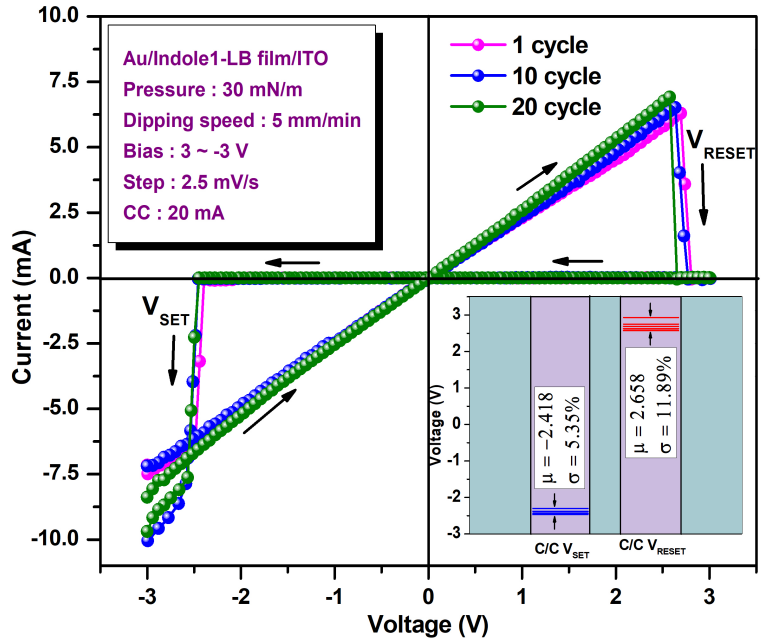

**Figure 6.** I – V measurements of the Au/Indole**1**-LB/ITO device over consecutive cycles. Inset represents the mean value (μ) and standard deviation (σ) of $V_{SET}$ and $V_{RESET}$ voltages

In order to check the stability of the designed device, I-V curve of the same device have been measured with passage of time. Interestingly almost reproducible switching behaviour have been observed even after 60 days of device fabrication. Cumulative standard deviations of $V_{SET}$ and $V_{RESET}$ measured with the passage of time have been found to be 9.59% and 20.60% respectively (figure S6 of supporting information). Device yield have been calculated by measuring I-V sweep characteristics for each devices prepared using two LB films with 5 × 6 array device structure (figure S1 of supporting information). Here each film forms 30 switchable independent cells. Switching characteristics were found within the fifty two cells out of total measured sixty cells resulting a device yield of the order of 86.7%. As a whole data retention, repeatability and stability studies as well as high device yield suggested that switching devices using Indole1 molecule may be a potential candidate for memory application and realization of organic electronics.

Several mechanism and strategies such as reduction-oxidation, electron tunnelling and hopping, ionic conduction, conformational change, space charge limited conduction and traps etc have been proposed to explain the observed switching behaviour.[22,42,45] The device having configuration Au/Indole**1**/ITO involves carrier conduction through injection and transport. Built-in internal field developed due to difference in work functions of the metal electrodes used and the amplitude as well as the direction of bias voltage, electron injection from the metal electrode to the LUMO level of the Indole1 molecules take place. This intern switches the device to its ON state via electro-reduction.[22] Conjugation within the molecules gets extended resulting lowering of HOMO-LUMO energy gap.[22] This leads the device to its high conducting (low resistance) ON state. The observed electro-reduction of Indole1 molecules may involve two possibility- (i) Injection of electron, when it overcomes the difference in energy between the LUMO level of the molecule and the work function of the ITO electrode. (ii) Transport of electron via applied bias dependent hopping.

Again depending on the bias condition when molecule gets oxidised, the device will return to its low conducting OFF state. Possible oxidation-reduction steps of Indole1 molecule have been shown in figure 7.

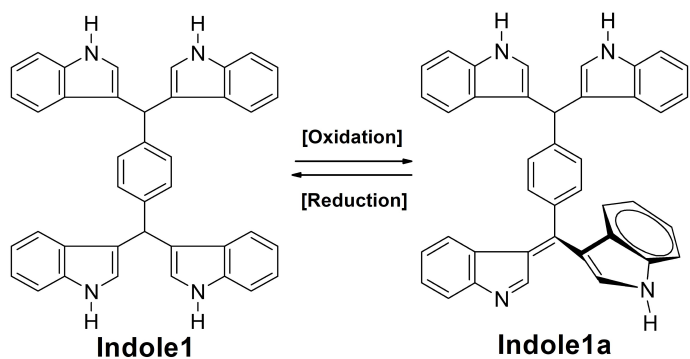

**Figure 7.** Oxidation - Reduction steps of **1**.

To confirm the role of reduction-oxidation process on the observed bipolar switching we have synthesized Indole2 molecule, where –H of Indole1 has been replaced by a methyl group and checked the I-V curves of the device using Indole2. Due to the presence of methyl group oxidation-reduction of the Indole2 molecule is hardly possible.[28] Interestingly here no bipolar switching was observed. Molecular structure of Indole2 and I-V curve for the device using Indole2 molecules has been shown in figure S7 of supporting information. This observation supports the role of reduction-oxidation process in the observed bipolar switching for the devices using Indole1 molecules.

Cyclic Voltammetry (CV) measurement also clearly supports the role of reduction-oxidation process on observed bipolar switching.[22,47,48] We have studied the CV curve for both the molecules Indole1 and Indole2 (Figure S8 of supporting information). Resulting CV curve of Indole1 molecules showed prominent oxidation and reduction peaks. On the other hand, the CV curve corresponding to Indole2 molecules did not show any prominent oxidation-reduction peaks.

In order to elucidate the conduction mechanism further dependence of the SET and RESET threshold voltages on the film thickness of the active layer of switching devices have been investigated. In order to do that switching devices have been constructed with LB film of various thickness such as 10, 60, 70, 80 and 90 layers and their I-V characteristics were measured. Corresponding results are shown in figure S9 of supporting information. Results suggested that switching voltage gradually increases from $V_{SET}$ = 1.06 to 3.51 V and $V_{RESET}$ = -0.879 to -3.56 V. Increase in switching voltages with active layer thickness indicates the possibility of electric field driven conduction in addition to oxidation – reduction mechanism.[51,52]

To understand the observed switching mechanism schematic of the energy level diagram under various conditions have been shown in figure 8. The slope of the LUMO energy level is 'after contact' condition determines the direction of electron transport from the ITO side under reverse bias condition while scanning from $+V_{max}$ to $-V_{max}$ (figure 8i).

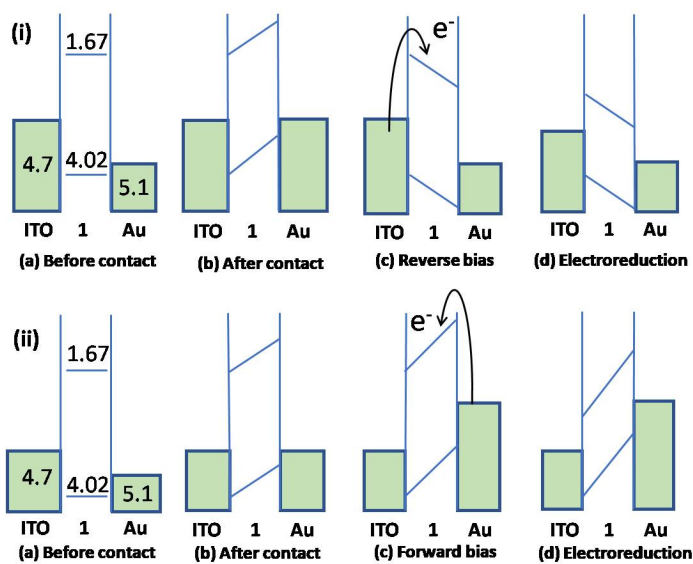

**Figure 8.** Energy level diagram of Au/Indole**1**-LB film/ITO configuration under (i) reverse bias and (ii) forward bias condition shown in four steps: (a) before contact with the electrodes, (b) after contact, (c) under bias, and (d) after electro-reduction of Indole1. Metal work functions and HOMO and LUMO values of Indole1 molecules are given in eV.

Hence in this device, a combined effect of barrier potential and electron injection due to bias induced reduction favours electron transport from ITO electrode to LUMO level compared to that from Au electrode under reverse bias condition. While scanning from $-V_{max}$ to $+V_{max}$ at threshold ($V_{RESET}$) oxidation of the Indole1 molecules occurred resulting in the device switched OFF. On the other hand, under forward bias condition while scanning from $-V_{max}$ to $+V_{max}$ at threshold voltage Au electrode donates an electron to Indole1 molecule within the device (figure 8ii). As a result, molecules near the Au electrode get reduced and the device switched to its ON state. Here due to higher barrier height electron reached to the LUMO level via hopping transport.[22] Again while scanning from $+V_{max}$ to $-V_{max}$ at particular threshold ($V_{RESET}$) oxidation of the molecules occurred and the device switched back to its OFF state.

To have further insight about the conduction mechanism of the Au/Indole1/ITO device, we analysed the I-V curves in double logarithmic scale as shown in figure 9. For both positive (figure 9a) as well as negative bias (figure 9b), the I-V curve shows linear behaviour under low voltage region. However, at higher voltage regions it follows quadratic behaviour.[42,46] Such charge transport behaviour can be explained in terms of one electron injected trap controlled Space Charge Limited Conduction (SCLC) theory.[46,49] At low voltage, the linear behaviour follows the Ohm's law due to thermally generated free-electron within the film. The quadratic behaviour at higher voltage region follows the Child's law where $I \propto V^2$.[42] To analyse further for the higher voltage quadratic region of the curve we have plotted I vs $V^2$ as shown in figure 9c and 9d. Almost linear nature of the curves under both bias condition suggested our assumption that here current-voltage relation obeys Child's law.[49]

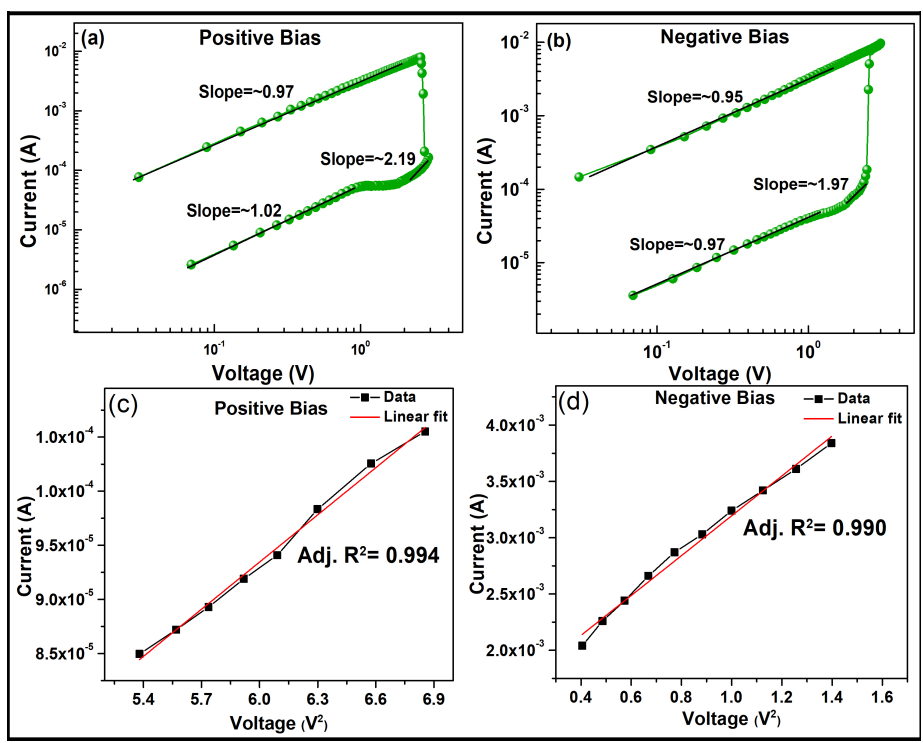

**Figure 9.** I–V characteristics (double logarithmic) of Au/Indole1-LB/ITO device in a) positive bias and c) negative bias. The plot of I vs $V^2$ (Child's law region) for the high slope part of the b) positive bias and d) negative bias.

**Conclusions**

In conclusion we have designed and investigated resistive switching devices based on organic molecule Indole1. LB and spin-coating techniques were used to prepare active layer of the device. Device fabricated using both the techniques showed bipolar resistive switching behaviour suitable for organic memory application under optimized measurement protocol. Bias dependent oxidation and reduction as well as electric field driven conduction mechanism were the key behind the observed switching. Analysis of I-V characteristics revealed that LB film based switching devices showed high memory window (~$10^6$) than that of spin-coated film based devices (~$10^4$). This difference in memory window may be due to the differencing active layer morphology prepared using two techniques as observed by AFM and SEM imaging studies. Data retention capability of the device was tested up to 5400 seconds with very good and satisfactory performance. Performance of the device was also found to be very good in terms of data reproducibility (more than 50 cycles), stability (more than 60 days) as well as device yield (~86.7%). As a whole our investigation revealed that switching device based on Indole**1** molecule may be a potential candidate for application in RRAM device using organic molecules.

**Supporting Information.** Schematic diagram for device Au/Organic material/ITO configuration (figure. S1), π – A isotherm of Indole1 showing value of limiting molecular area (figure. S2), dimension of Indole**1** (figure.S3), BAM image of air-water interface (figure. S4), Bipolar resistive switching of Au/Indole1-spin coated/ITO device (figure. S5), I – V characteristics study for stability of the device (figure.S6), Chemical structure of Indole2 (figure. S7) and CV of Indole1 & Indole2 (figure. S8).

**ORCID**
Syed Arshad Hussain: 0000-0002-3298-6260

**Author Contributions**
S.A.H. designed the work. S.S. and H.B.performed all experiments and data analysis. S.M. and B.D.synthesized the material. S.A.H. and S.S. wrote the manuscript with input from S.M., D.B. andP.K.P.

**Acknowledgments**
The authors are grateful to DST, for financial support to carry out this research work through FIST – DST project ref. SR/FST/PSI-191/2014. SAH is grateful to DST, for financial support to carry out this research work through DST, Govt. of India project ref. No. EMR/2014/000234. The authors are grateful to Department of Science and Technology (DST), Govt. of India for providing a 400 MHz NMR facility under DST-FIST programme (No SR/FST/CSI-263/2015). The authors are also grateful to UGC, Govt. of India for financial support to carry out this research work through financial assistance under UGC – SAP program 2016. The authors acknowledge the central instrumentation centre (CIC), Tripura University for providing AFM and SEM facility.